\documentclass{aastex6}

\begin{document}          


\title{The Mass of the Cepheid V350 Sgr\altaffilmark{1}} 

\author{Nancy Remage Evans,\altaffilmark{2}
Charles Proffitt,\altaffilmark{3}
Kenneth G. Carpenter,\altaffilmark{4}
Elaine M. Winston,\altaffilmark{2}
Gladys V. Kober,\altaffilmark{4} 
H. Moritz G\"unther,\altaffilmark{5}
Natalia Gorynya,\altaffilmark{6,7} 
Alexey Rastorguev,\altaffilmark{7,8}
and L. Inno, \altaffilmark{9}
 }

\altaffiltext{1}
{Based on observations with the NASA/ESA {\it Hubble Space Telescope\/} obtained
at the Space Telescope Science Institute, which is operated by the Association
of Universities for Research in Astronomy, Inc., under NASA contract
NAS5-26555.}

\altaffiltext{2}
{Smithsonian Astrophysical Observatory, MS 4, 60 Garden St., Cambridge, MA
02138, USA; nevans@cfa.harvard.edu}


\altaffiltext{3}
{Space Telescope Science Institute, 3700 San Martin Drive, Baltimore, MD
21218, USA}

\altaffiltext{4}
{NASA Goddard Space Flight Center, Code 667, Greenbelt, MD 20771, USA}

\altaffiltext{5}
{Massachusetts Institute of Technology, Kavli Institute for Astrophysics and Space Research,
77 Massachusetts Avenue, NE83-569, Cambridge, MA 02139, USA}

\altaffiltext{6}{Institute of Astronomy, Russian Academy of Sciences, 48 Pyatnitskaya str.,
Moscow, 119017, Russia}

\altaffiltext{7}
{Sternberg Astronomical Institute, Lomonosov Moscow State University, 13 
Universitetskij prosp., Moscow, 119991, Russia}

\altaffiltext{8}
{Faculty of Physics, Lomonosov Moscow State University, 1, bld. 2, Leninskie
gory, Moscow, 119991, Russia}


\altaffiltext{9}
{Max Planck Institute for Astronomy in Heidelberg}





\begin{abstract}

V350 Sgr is a classical Cepheid suitable for mass determination.
It has a hot companion which is prominent in the ultraviolet and which is 
not itself a binary.  We have obtained two high resolution echelle 
spectra of the companion at orbital velocity maximum and minimum with 
the Hubble Space Telescope (HST) Space Telescope Imaging Spectrograph
(STIS) in the 1320 to 1510 \AA\/ region.  
By cross-correlating these spectra we obtained 
the orbital velocity amplitude of the companion with an uncertainty in the 
companion amplitude of 1.9 km sec$^{-1}$. 
This provides a mass ratio of the Cepheid to the companion of 2.1.  
The ultraviolet energy distribution of the companion provides
the mass of the companion, yielding a 
 Cepheid mass of 5.2 $\pm$ 0.3   M$_\odot$.
This mass requires
some combination of moderate main sequence core convective overshoot 
and rotation to match evolutionary tracks. 

\end{abstract}

 
\keywords{stars: variables: Cepheids; (stars:) binaries: spectroscopic; 
stars: fundamental parameters  }


\section{Introduction}

Masses are the most fundamental parameter governing the evolution
of single stars. Interactions between stars in binary/multiple systems can, 
of course, alter a mass in interesting ways.  The tension between the masses 
derived from evolutionary calculations and those from pulsation calculations
has been designated ``the Cepheid mass problem''.  A good summary of
 the differences and the recent state is provided in Neilson, et al 
(2011), who conclude that it still exists at the 10-20\% level. 
This  implies uncertainty in the evolutionary predictions of luminosity 
for post-main sequence He burning stars.   For classical Cepheids 
evolutionary calculations are also
important in making any adjustments needed to the Leavitt (Period-Luminosity) Law
for differences in metallicity between galaxies.

Observed masses are needed to clarify these questions.  In the Milky Way (MW) 
there are no Cepheids known in eclipsing binaries.  The advent high resolution 
spectra in the ultraviolet (UV) from satellite observations
(originally the International
Ultraviolet Explorer [IUE] and currently the Hubble Space Telescope [HST])
has provided orbital velocity amplitudes of the hot companions of several 
Cepheids.  Combining this amplitude with the ground-based orbital velocity amplitude 
for the Cepheid, and a mass of the companion from the energy distribution 
in the ultraviolet provides a Cepheid mass.   In addition, a  dynamical mass
has been determined for Polaris using HST astrometry (Evans, et al. 2008, 2018). 
An upper limit to the 
 mass for W Sgr has been derived from HST astrometry.  
A summary of the references
and results is provided by Evans, et al. (2011).  
In several cases improved masses are anticipated 
soon, largely because of the incorporation of interferometry to provide 
additional resolved orbits.  The first result of this program is V1334 Cyg 
(Gallenne, et al. 2018).

Not only is the determination of  masses in the MW improving,  an additional
valuable comparison has become possible with Cepheids in 
the  Large Magellanic Cloud (LMC).  
Several  eclipsing binaries
have been discovered in the  LMC
(Gieren, et al. 2015; Pilecki, et al. 2013; Pilecki, et al. 2015; summarized by
Pilecki et al 2018). 
  Thus a comparison
of the mass luminosity relation can be made at two metallicities.

The first step in mass determination is the derivation of a binary 
orbit for the primary (Cepheid) from ground-based spectra, which is
available for many stars.  An early result from  UV studies of the companions
is that a substantial fraction of the companions are themselves binaries
(e.g. Evans, et al. 2005). This is to be expected in high and intermediate
mass systems, but the additional observations needed to determine a mass
are often prohibitively expensive of telescope time.  

The system containing the Cepheid V350 Sgr = HD 173297 is one where 
previous UV observations found the companion to be single. It was observed 
twice with the HST Goddard High Resolution Spectrograph (GHRS) 
 medium resolution (R = $\lambda$/$\Delta\lambda$ $\approx$ 20,000) between 
1840 and 1880 \AA\/ in 1995 (Evans, et al. 1997).  From the velocity
difference between phases of these two spectra, and the velocity 
difference from the Cepheid orbit, they derived a mass ratio 
$M_{Cep}/M_{Comp}$ = 2.1 $\pm$ 0.3.  Using the mass from the UV energy 
distribution of the companion (B9.0 V; Evans and Sugars 1997), they derived
a mass for the Cepheid of 5.2 $\pm$ 0.9  M$_\odot$.  

Since that discussion, a number of factors have contributed to an improved 
analysis of the system.  A new orbit has been derived based on considerable
additional velocity data (Evans, et al. 2011),   particularly including  
data near minimum velocity.  Because the orbital period is very close
to 4 years, uniform phase coverage has been difficult to obtain.  
In the project here HST spectra obtained with the Space Telescope
Imaging Spectrograph (STIS) are combined with this new orbit, 
providing improved velocities of the companion
as discussed in the next section.  Successive sections below discuss 
the observations, 
 the details of velocity measurement from these spectra, 
the companion, and 
the results and implications of the new measurements.   

{\bf Gaia will, of course, ultimately be important in improving 
the distance and mass.  
However, the current DR2 release does not include binary motion in the solution.
To illustrate, the expected parallax based on the distance
from the Benedict et al. (2007)
Leavitt law is 1.12 mas.  The semi-major axis of the orbit (Evans, et al. 2011)
a sin i is 1.32 AU, which is 1.48 mas at this distance.  Hence the Gaia solution
including orbital motion is clearly needed.
Because of this, and also concerns about the effect of Cepheid light 
variation and possibly the brightness of the system, the appropriate solution
will come with later Gaia releases.  
}



\section{Observations}

STIS observations were obtained in HST Cycles 21 and 23, sampling 
orbital velocity maximum and minimum.   The STIS spectra provide
several improvements over the GHRS spectra.  The high resolution 
echelle mode was used (E140H resolution 114,000 in place of 
the GHRS G200M resolution of 20,000).  The much larger wavelength 
range (1320 to 1510 \AA) provided many more spectral lines for 
the velocity measurement.  Finally, the new orbit resulted in 
phases selected for better optimization of the velocity amplitude
measurement.   
The observations are summarized in Table~\ref{exp.date}, which includes
the orbital velocity of the Cepheid at the time of the observations.

\begin{figure}
\includegraphics[width=12 cm,angle=0]{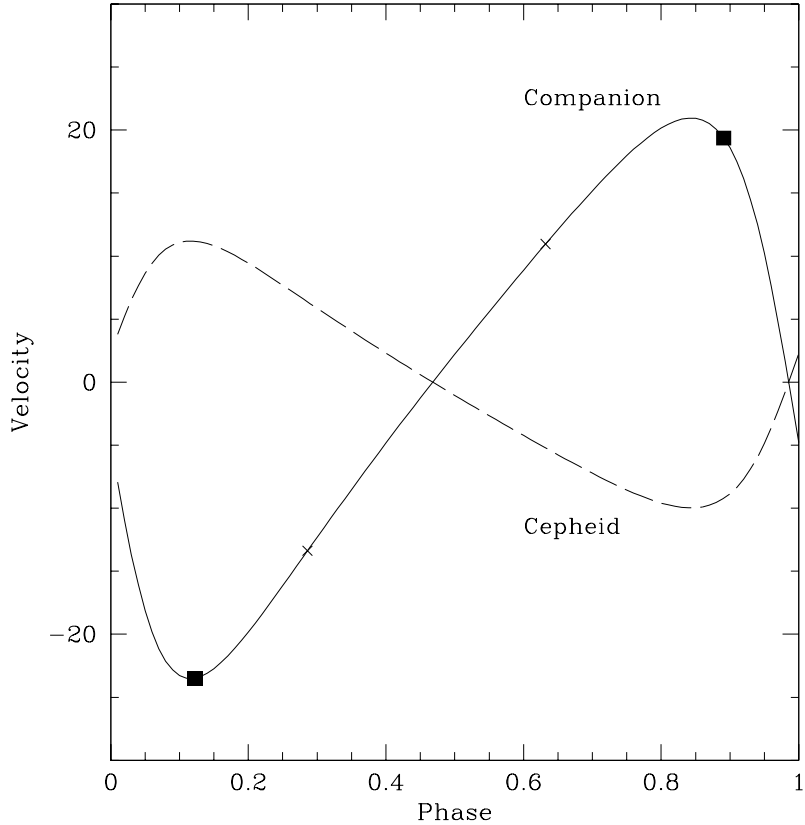}
\caption{{\bf Schematic summary of the phases of observations 
of V350 Sgr B.  Dashed line: Cepheid
orbital velocity curve; solid line: companion orbital velocity 
curve; filled squares: phases of STIS observations; 
x's: phases of previous GHRS 
observations. Velocities are from the orbit, not measured 
velocities as discussed in the text.
Velocities are in km sec$^{-1}$} 
  \label{phas}}
\end{figure}

{\bf To sumarize the observations we provide the schematic in Fig~\ref{phas}.  
The Cepheid orbit is from Evans, et al. 2011; the orbit of the companion 
anticipates the result of this paper for the orbital velocity ratio.  
The phases of the STIS observations and the previous GHRS observations are 
shown.  For the best velocity accuracy, we have cross-correlated the two 
STIS spectra to derive the velocity difference between the two, 
rather than using a template of a different star to determine the velocity
of each individual star.  Previous experience has shown that cross-correlation
between two observations of the same star produces much better defined
results, particularly for weak spectra, 
since the lines have the the same abundances, rotation velocity, 
and microturbulence.  
Thus,  Fig~\ref{phas} is a schematic  
to indicate the phases of the observations, but not measured velocities.

Using these observations, the reductions were done as described in the next section. 
The velocity difference measured for the companion was compared with that 
of the Cepheid at the same phases to determine the mass ratio between the 
two stars.  To determine the mass of the Cepheid, the mass of the companion is
needed.  Since it is on the main sequence, the companion mass 
is comparatively well known, 
and an ultraviolet spectrum provides an energy distribution.  
The determination of the companion mass 
now includes both updated masses from eclipsing binaries
(Torres, et al. 2010), as well as comparison with recent  model atmospheres
(Bohlin, et al. 2017).  To summarize, comparisons were made with other Cepheid
masses in the MW and the LMC.
}

\begin{deluxetable}{lllll}
\tabletypesize{\footnotesize}
\tablecaption{STIS Observations of V350 Sgr\label{exp.date}}
\tablewidth{0pt}
\tablehead{
\colhead{Year\/\/\/\/} & \colhead{Dates\/\/\/\/} & \colhead{\/\/JD\/\/\/}  
 & \colhead{$\phi_{orb}$\/\/\/\/}   & \colhead{V$_r$$_{orb Cep}$\/\/\/\/} \\
\colhead{ } & \colhead{} & \colhead{-2,400,000}  
 & \colhead{ }   & \colhead{km s$^{-1}$} \\
}
\startdata
2013 & Oct. 1-5 (Oct 3) &  56599  & 0.123  &  11.1  \\
2016  & Aug. 23-28 (Aug 25) &  57625 & 0.819  &  -9.9 \\
\enddata
\end{deluxetable}

\section{Reductions}

The two sets of observations of V350 Sgr each consist of 12
individual spectra, which  each  contain 40 echelle orders, covering
 1320\AA\/ to 1510\AA.
  The flux calibration of the STIS echelle gratings is sensitive 
to changes in the alignment of the echelle blaze function (Bowers and Lindler 
2003). Misregistration of these blaze functions can impose artificial patterns 
on the calibrated flux for each spectral order. This is particularly problematic 
when attempting to cross correlate spectra taken at different times, since for 
broad-lined stars the change in spectral shape resulting from the misaligned 
blaze can be difficult to cleanly separate from a velocity shift.

The STIS calibration pipeline applies a correction for the expected blaze 
function shift, which depends on both the wavelength offset measured in the 
contemporaneous lamp calibration spectra and on the date of the observation 
(see Alosi 2011). The latter term is needed because the blaze function shift 
for the STIS echelle gratings has been shown to evolve systematically over 
time. Unfortunately, at the time the bulk of our analysis was done, the 
time dependent terms for the post-SM4 blaze function shifts were not yet 
available in the default pipeline calibration. This led to very large 
misalignments in the flux calibration. Flux inconsistencies in the overlap 
between spectral orders of 10\% or more were common.

To correct for the blaze function misalignment, we developed a simple IDL 
script which recovers the applied sensitivity curves for each spectral order 
from the net and flux vectors delivered in the pipeline calibrated spectra, 
and then finds the overall shift of those sensitivity curves which makes 
the calibrated flux in the 
wavelength overlap between spectral orders most consistent.
The approach we used is a preliminary version of the algorithm
in Baer, Proffitt, and Lockwood (2018).

Once this correction has been applied, it is still necessary to combine 
the multiple, rather faint observations, each containing 40 different 
echelle orders, into a single 1D spectrum for each of our two epochs.
 We first define our final output wavelength grid, which is chosen to 
have the same average dispersion as our echelle observations, but with 
uniform spacing in delta(log lambda). For each individual observation, 
we then interpolate the flux and error at each wavelength bin onto this 
output grid. Where two echelle orders overlap in wavelength, we weight 
their contributions at each wavelength by the relative sensitivities at 
that wavelength. We then combine all of the separate observations, this 
time weighting by their relative exposure times. This simple interpolation 
procedures does introduce some smoothing and results in some correlations 
between adjacent wavelength bins which are not properly taken into account 
by simply interpolating the error vector. However, for our science goals, 
preserving the spectral resolution and the absolute wavelength calibration 
are the highest priority, and as a practical matter, this combined spectrum 
will normally be further smoothed prior to cross-correlation to determine 
the velocity difference.  We estimate that the mean S/N per resolution element
is approximately 10.  

Note that attempting to combine these echelle spectra by weighting using 
the pipeline estimated errors instead of the relative throughputs or exposure 
time results in biasing the coaddition towards data points that happen to 
have fluctuated low, and for relatively low S/N data such as we obtained 
for V350 Sgr, this can lead to significant problems in the spectral combination 
and cross-correlation. 

\begin{figure}
\includegraphics[width=12 cm,angle=0]{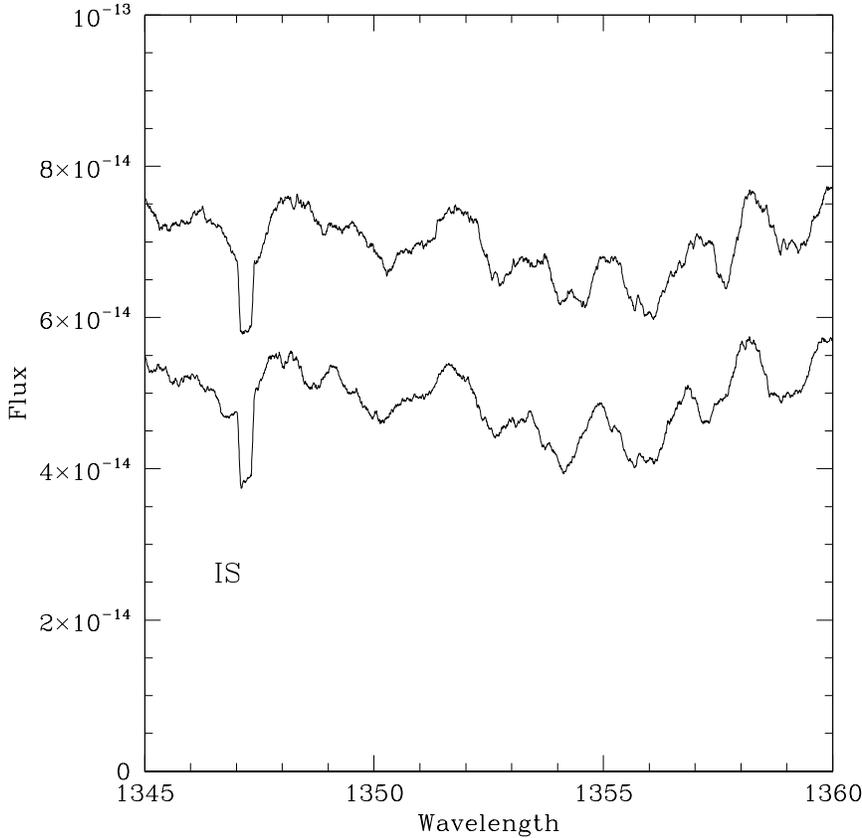}
\caption{An illustrative portion of the spectra of the 2016 (top) and 
2013 (bottom) observations after coaddition.  The sharp feature at 1347 \AA\/ is an 
interstellar line (slightly broadened by coaddition and smoothing), 
 It is easily identified and removed by interpolation.  
Wavelength is in \AA; 
flux is in ergs sec$^{-1}$.  The top spectrum has been offset for clarity
  \label{spect}}
\end{figure}

{\bf Representative regions of the spectra are shown in Fig~\ref{spect}.}

The summed spectra for each epoch were then ``blemished'', 
to remove interstellar medium (ISM) features, easily recognizable because they are 
narrow compared with the stellar lines.  This included several $^{12}$CO 
lines, again recognizably narrow.  



The spectra at the 2 epochs were cross-correlated using an IDL routine.  
The resulting spectra are still relatively weak, 
however the single
quantity--the radial velocity difference between the two--can still be 
determined relatively well from two spectra with identical properties:
temperature, abundance, and rotation.   In measuring the velocity, we 
explored a number of parameters.  The cross-correlations were done 
in a series of pieces, typically 10 \AA\/ wide.  However, the boundaries
of these regions were carefully selected so that the relatively broad lines
are contained within a region (not sliced in half).  
The effect of smoothing was 
explored, again since the stellar features are broad.  The final 
smoothing was selected based on reduced noise in the broad stellar
features (20 point smooth which corresponds to approximately 0.11 \AA).
A gaussian was fitted to the cross-correlation result, with particular
attention paid to the background identification in the fitting. 

A series of tests was also performed using several IDL cross-correlation
routines, data treatments such as smoothing, and Gaussian fitting 
widths.  These involved typically cross-correlation of the whole 
spectrum, and also a truncated version omitting a prominent feature
which was contaminated by strong ISM absorption.  








\subsection{Velocity}

The  individual cross correlations of pieces of the spectra were inspected, 
and three were removed because the gaussians were unsatisfactory.
Three further segments were removed as outliers.  
The remaining reliable values of the velocity difference from 
13 segments is -43.3 $\pm$ 1.9 km sec$^{-1}$, which we use as the estimate
for the uncertainty of the velocity.  Considering the rotation 
velocity of the star (75-100 km/s), this is a reasonable uncertainty. Similar 
treatment of a more highly smoothed spectrum (50 point smooth)
provided a velocity within these errors.  The results of the tests 
using the full spectrum and the truncated version were consistent 
with this velocity and error estimate.  





The velocity uncertainty is dominated by the number  of broad features
from which velocities can be measured.   STIS wavelength precision  is 
very high using standard observing procedures, such as peakup during 
acquisition.  Ayres (2010) has confirmed this, and our estimated 
instrumental error is 0.5 pixel corresponding to 0.75 km sec$^{-1}$, 
which is only a small contribution to the error.   
We have confirmed this using the repeatability of the ISM lines of 
two other Cepheids with similar observations (S Mus and V1334 Cyg). 
The RMS scatter in about the mean velocity is only 0.24  km sec$^{-1}$
(Proffitt et al. 2017).


\section{The Companion}
The mass of the companion must be determined to complete the determination 
of the Cepheid mass.  It is based on an International Ultraviolet 
Explorer (IUE) spectrum, as discussed by Evans and Sugars (1997). 
An important aspect of this approach to mass determination is that 
the spectrum of a late B companion is completely uncontaminated 
by the brighter Cepheid for wavelengths shorter than 1700 \AA.  
In this spectral type range the situation is favorable since the 
energy distribution is very temperature sensitive and mass changes
comparably slowly as a function of spectral type.  

We add to the 
discussion in Evans and Sugars  in two ways using 
temperatures determined from model atmospheres and using a 
recent list of masses determined from eclipsing binaries.  

To assess the energy distribution, the spectrum must, of course,
be corrected for reddening.  As in previous discussions (Evans 1991),
the reddening is derived from optical colors which have been 
corrected for the comparatively small effects of the light from 
the companion, resulting in E(B-V) = 0.32 mag in the case of V350 Sgr. 
This is then corrected to the E(B-V) which would be seen by 
the broadband colors of an OB 
star [E(B-V) = 0.36 mag] before applying the reddening law of 
Cardelli, Clayton, and Mathis (1989)  to the spectrum  using the 
IDL routine UNRED\_CCM.  Because the star is relatively close and 
the extinction is low, this extinction law should be applicable.

\begin{figure}
\includegraphics[width=12 cm,angle=0]{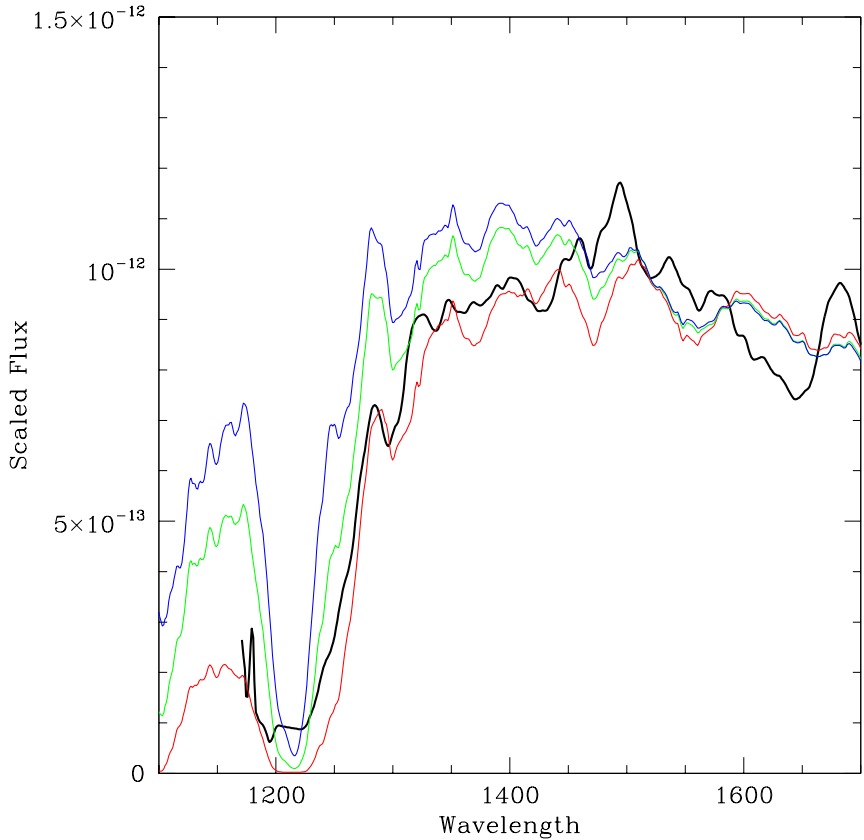}
\caption{Comparison of the IUE spectrum of V350 Sgr B with 
model atmospheres. Lines show:
 black:  V350 Sgr B; red: 10750 K; green:  12000K; and 
blue: 13000 K.  The IUE spectrum and the atmospheres have   all been 
smoothed to emphasize the energy distribution. Wavelength is in \AA; 
flux is in ergs sec$^{-1}$.
  \label{sed}}
\end{figure}

The analysis is focused on the temperature sensitivity of the 
energy distribution between 1150 and 1700 \AA.  
The spectrum was compared with ``BOSZ'' model 
atmospheres from Bohlin, et al. (2017) for 
temperatures 10750 K, 12000 K, and 
13000 K (log g = 4.00, turbulence = 2 km sec$^{-1}$)
corresponding closely to B9 V, B8 V, and B7 V respectively.  The models and the 
IUE spectrum have been scaled to the wavelength region  1500 to 
1700 \AA\/ for comparison (Fig~\ref{sed}).   
Fig~\ref{sed} shows 
that the V350 Sgr B is slightly hotter than 10750 K, but distinctly 
cooler than 12000 K.   To make the comparison more quantitative, the ratio of 
the flux in bins from 1250 to 1350 \AA\/ to the flux from 1500 to 1700\AA\/ 
was computed (Table 2).  Using these ratios, the temperature of V350 Sgr B
is estimated to be 11000 K.  

Ultraviolet spectra are particularly sensitive to interstellar extinction. 
However, this is not as serious a contributor to the estimate of the mass 
of the companion as might be expected.  The E(B-V) of Cepheids has been well 
studied, and the value for the  V350 Sgr system is moderate.  Care has been 
taken in the correction of the energy distribution for reddening (above). 
In particular, the energy distributions of late B stars are very temperature 
sensitive in the region of Fig~\ref{sed}, and the temperature is estimated from 
a relatively short wavelength range (Table 2).  To illustrate that 
a moderate uncertainty in the reddening does not distort the interpretation 
of the ultraviolet spectrum (Fig~\ref{sed}), the bottom two lines in Table 2, 
show the range of the flux ratio for the range of E(B-V) = 0.33 to 0.39 mag. 
The flux ratios remain between B9 V and B8V.     

\begin{deluxetable}{llll}
\tabletypesize{\footnotesize}
\tablecaption{Spectral Comparisons \label{spec.comp}}
\tablewidth{0pt}
\tablehead{
\colhead{Spectrum } & \colhead{Flux Ratio} & \colhead{ E(B-V)} &   \colhead{ M} \\ 
\colhead{Type} & \colhead{1300/1600 } &  \colhead{ mag } &  \colhead{ M$_\odot$ }  \\
}
\startdata
V350 Sgr B  & 0.781 & 0.36 &  \\
B9 V 10750 K & 0.707  & & 2.41 \\
B8 V 12000 K & 0.928 & & 2.78 \\
B7 V 13000 K & 1.044 & &  3.22 \\ 
  & & & \\
V350 Sgr B  & 0.818 & 0.39 & \\
V350 Sgr B & 0.748 & 0.33 & \\
\enddata
\end{deluxetable}

\begin{figure}
\includegraphics[width=12 cm,angle=0]{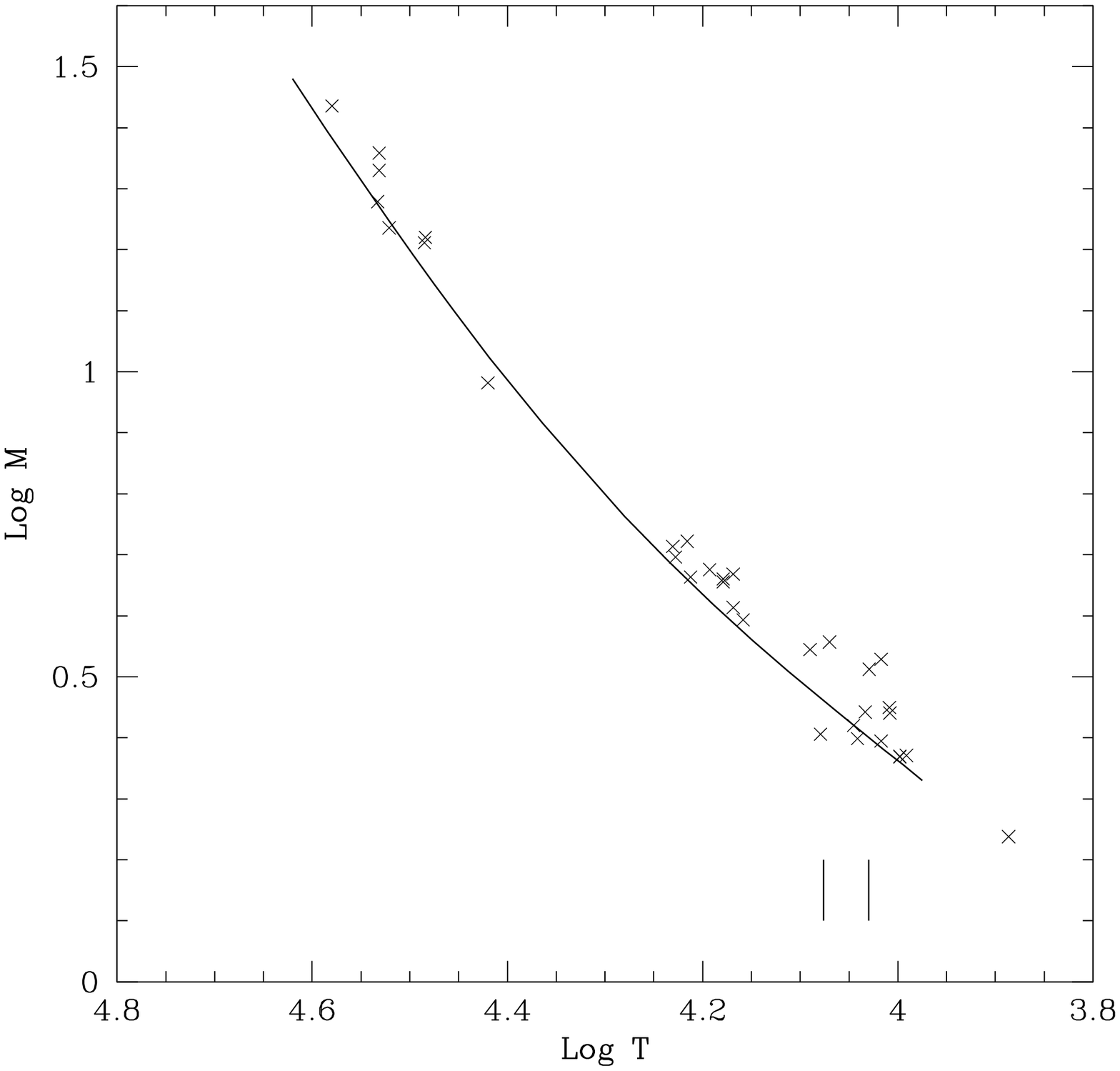}
\caption{Masses and temperatures of eclipsing binaries.
Masses and temperatures from 
Torres et al.: x's;  Harmanec (1988)
relation (0.02 lower in log mass to provide
a lower envelope): line; small vertical lines near the bottom show the 
temperatures of the B8V and B9V models.  Masses are in  M$_\odot$;
temperatures are in K.
  \label{mass}}
\end{figure}

In order to determine the mass corresponding to this temperature, 
we have used the data of Torres, et al (2010).  Fig~\ref{mass} shows 
 their temperature and mass data for O and B stars.  For comparison 
 with the discussion of Evans and Sugars (1997)
we use the mass-temperature relation from Harmanec (1988). 
An important fact in assessing the mass of V350 Sgr B is that 
the age of the star is known, since it is a companion of a 
young Cepheid.  This means that it will lie very close to the 
zero age main sequence, in contrast to many of the eclipsing 
binaries which will have evolved significantly beyond. 
The spread in ages for the eclipsing binaries is responsible 
for much of the scatter in Fig~\ref{mass}.  For 
this reason, it is the lower envelope in Fig~\ref{mass}  which is 
appropriate for our comparison.  
The Harmanec relation lowered by 0.02 in log mass
provides a good lower envelope. Fig~\ref{sed}  and Table 2 show that V350 Sgr B 
is slightly warmer than the B9 V model (10750 K), but cooler than the 
mid-point between B9 V and B8 V (which we will call B8.5 V).  Using 
the masses in Table 2 (Harmanec envelope), the mass between B9 V and
B8.5 V is 2.50 M$_\odot$, which is 0.1  M$_\odot$ from either 
 B9 V or B8.5 V, which are ruled out by Fig~\ref{sed}.  This is the same as 
the result from Evans and Sugars using MK spectral classes.

\section{Results}

The orbital velocity difference between the observed phases (very close
to the orbital velocity  amplitude) of V350 Sgr B (-43.3 $\pm$ 1.9 km sec$^{-1}$)
can be compared with the orbital 
velocity difference for the Cepheid V350 Sgr A.  We have used the orbit of 
Evans, et al. (2011) to determine the velocity difference of the Cepheid 
at the same phases as the STIS observations, which is  21.0 km sec$^{-1}$
(Table 1). Thus the mass ratio M$_{Cep}$/M$_{Com}$ is  2.1$\pm$ {\bf 0.084}.  
Combining this with the mass of the companion  from the previous section
2.5 $\pm$ 0.1  M$_\odot$ (4\%)
 yields Cepheid mass of 5.2 M$_\odot$ with an error estimated from
the combined error of the velocity and companion mass of 6\%.








\begin{figure}
\includegraphics[width=10 cm,angle=0]{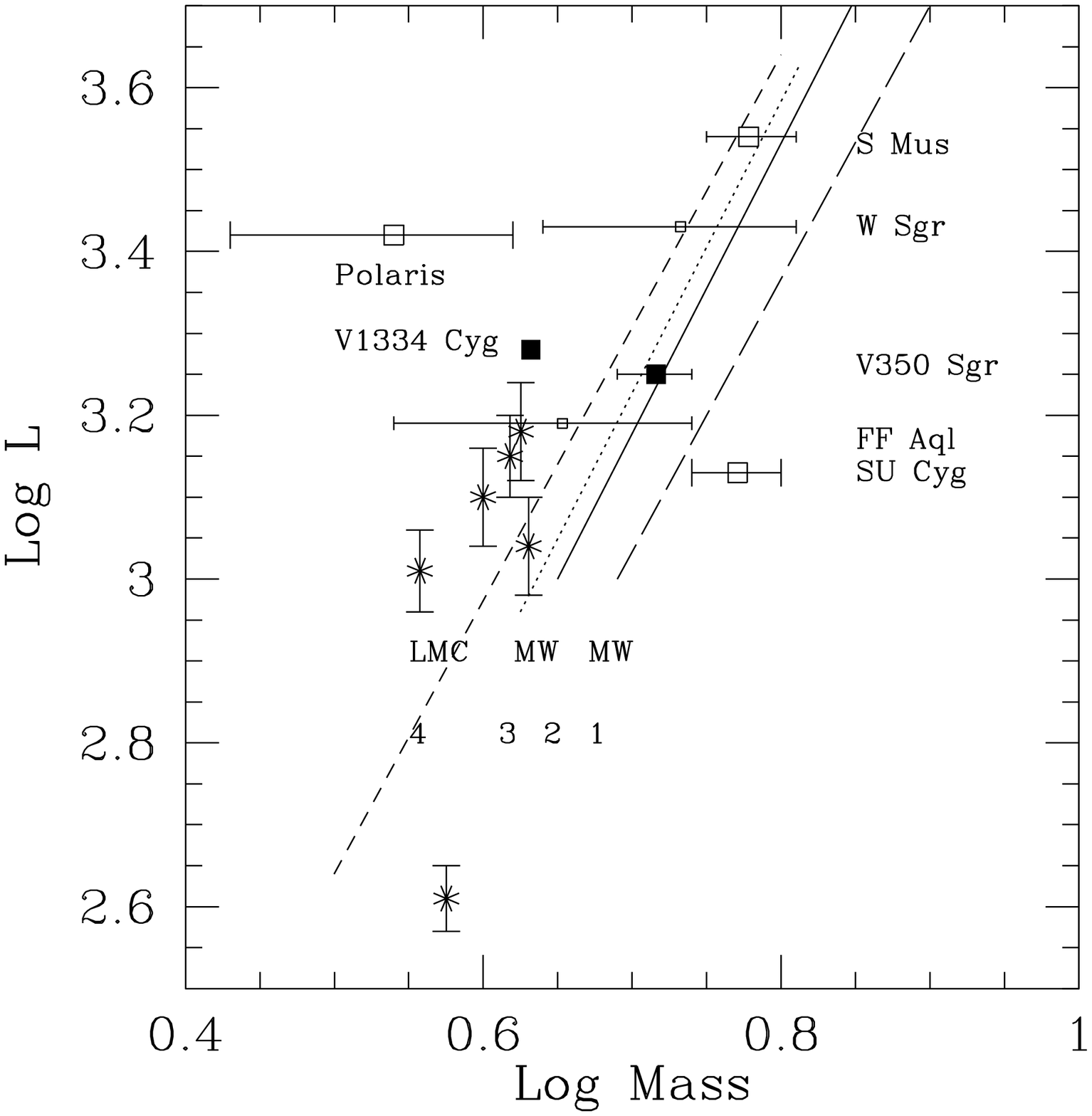}
\caption{Cepheid mass--luminosity (ML) relations.  {\bf MW Cepheid masses:}
 squares;
 V350 Sgr and V1334 Cyg: large filled squares; 
masses expected to be updated: large open 
squares; FGS astrometry incorporated: small open squares;
{\bf LMC Cepheid masses:}
asterisks; {\bf ML relations} from right to left (1, 2, 3, 4): 
1. no main sequence convective
overshoot, MW metallicity (Bono, et al. 2016): 
long dash line; 2.  moderate  convective
overshoot, MW metallicity (Bono, et al. 2016): solid line; 
3. small  convective
overshoot, and rotation, MW metallicity (Anderson et al. 2014): dotted line; 
4. LMC metallicity, moderate convective overshoot (Bono, et al. 2016): 
short dash line. Since tracks for 4 M$_\odot$ do not become hot enough to 
enter the instability strip on the second crossing, most of the relations 
have been truncated at lower masses.  
However, for comparison with the LMC first 
crossing star, the LMC relation has been extended 
to lower luminosities.  See text
for discussion.
  \label{ml}}
\end{figure}


\section{Discussion}

Fig~\ref{ml} puts the mass of V350 Sgr in the context of the measured Cepheid 
masses, and also of theoretical predictions. 
New masses are available for V350 Sgr (this paper), V1334 Cyg (Gallenne, et al.
2018) and Polaris (Evans, et al. 2018).  Other masses 
 for MW Cepheids are 
from the sources listed in Evans, et al. (2011). 
(Note that the mass for 
SU Cyg is a lower limit, and that of W Sgr is an upper limit.) W Sgr and 
FF Aql incorporate HST Fine Guidance System (FGS) astrometry 
(Benedict, et al 2007; Evans, et al. 2009).  
This is a ``before'' picture, 
since the  accuracy of 
the masses of S Mus and SU Cyg will be improved in near future including the 
results from interferometry (which will produce an  ``after'' picture).  
The mass for Polaris is preliminary, and will ultimately 
be improved using  CHARA interferometry, 
but because of the long period of its
orbit, this will not be for several years.

Luminosities in Fig~\ref{ml} 
for the MW Cepheids are derived from the 
Leavitt Law (Period-Luminosity) of 
Benedict, et al. (2007).  Alternately for V350 Sgr, a radius was derived 
using the modified Balona technique (Rastorguev and Dambis 2011)
after carefully correcting for the effect of the companion on the
light curve. The resulting 
luminosity is slightly smaller than that 
in Fig~\ref{ml}.  

The LMC Cepheids in eclipsing binaries have recently been 
rediscussed by Pilecki, et al (2018) and their masses and
luminosities are shown in Fig~\ref{ml} .  
This includes the interesting case of
LMC-CEP-1812 which is crossing the instability strip for the 
first time (the least luminous Cepheid 
in Fig~\ref{ml}) and may be 
a merger product (Neilson et al. 2015).  It occurs approximately 
0.2 in log(L/L$_\odot$)  lower than the relation in 
Fig~\ref{ml} for second and 
third crossing stars 
as expected the predictions of Bono et al. (2016) from the comparison
of luminosities between crossings. 
In addition the system LMC-CEP-1718 A and B 
contains a pair of first overtone pulsators (the two most massive 
LMC stars in Fig~\ref{ml}). 
The combination is 
 unusual in that the more massive is less luminous. 
  However this may be explained by the uncertainty in the luminosities.


The range of theoretical predictions from evolutionary tracks is also shown.  The left 
(short dash) line is for the metallicity of the LMC; others are for MW metallicity.
The right hand line (long dash) shows the prediction for stars with no core 
convective overshoot on the main sequence (Bono, et al. 2016). As is well known, 
these predictions produce the lowest luminosity for a given mass. 
The two lines in the middle illustrate combinations of parameters which can 
increase the luminosity for a given mass by increasing the size of the 
central He core after core 
hydrogen burning.  The solid line has moderate convective
overshoot added (d$_{over}$ = 0.2 H$_p$ where H$_p$ is the pressure scale height).
The dotted line shows recent Geneva calculations (Anderson, et al. 2014) which 
include both a smaller amount of overshoot (d$_{over}$ = 0.1 H$_p$) and rotation.  
The value of 0.5 $\omega_{crit}$ (critical velocity)
actually represents the effects of a wide 
range of rotations well.  All the predictions 
in Fig~\ref{ml} are for combined second
and third crossings of the instability strip.

The improved accuracy of the mass of V350 Sgr confirms that 
evolutionary tracks without rotation or overshoot predict too low
a luminosity for the mass, which is in agreement with other masses in 
Fig~\ref{ml}.  
As improved masses become available, other parameters 
influencing the luminosity in the Cepheid stage will be more 
tightly constrained.

\noindent {\em Acknowledgments}: {\bf The referee's report improved
the clarity of presentation.}
Support was provided by HST Grant 
GO-13368.01-A and Chandra X-ray Center 
NASA Contract NAS8-03060 (to NRE) and HST-GO-13368.008-A (to CP) . 
A. Rastorguev acknowledges Russian Foundation 
for Basic Research (RFBR grant 18-02-00890) for partial support.


\end{document}